\documentclass[
 reprint,
 amsmath,amssymb,
 aps,
]{revtex4}
\usepackage{amsmath}
\usepackage{graphicx}
\usepackage{bm}
\usepackage[colorlinks]{hyperref}
\usepackage[mathlines]{lineno}
\usepackage[colorlinks]{hyperref}
\usepackage{mathrsfs}
\usepackage{xcolor}
\RequirePackage{graphicx}

\usepackage[
text={7in,10in},centering,
a4paper
]{geometry}

\begin{document}

\preprint{APS/123-QED}

\title{Coexistent physics of  massive black holes in the phase transitions}

\author{Ming Zhang}
 \email{mingzhang@mail.bnu.edu.cn}
\author{Wen-Biao Liu}
 \email{wbliu@bnu.edu.cn}
\affiliation{Department of Physics, Institute of Theoretical Physics,  Beijing Normal University, Beijing 100875, China}

\date{\today}

\begin{abstract}
The coexistent physics of de Rham-Gabada-dze-Tolley (dRGT) massive black holes and holographic massive black holes is investigated in the extended phase space where the cosmological constant is viewed as pressure. Van der Waals like phase transitions are found for both of them. Coexistent curves of reduced pressure and reduced temperature are found to be different from that of RN-AdS black holes. Coexistent curves of reduced Gibbs free energy and reduced pressure show that Gibbs free energy in the canonical ensemble decreases monotonically with the increasing pressure. The concept number density is introduced to study the coexistent physics. It is uncovered that with the increasing pressure, the number densities of  small black holes (SBHs) and large black holes (LBHs) change monotonically in the contrary directions till finally reaching the same value at the critical points of the phase transitions. In other words, with the increasing pressure the number density differences between SBHs and LBHs decrease monotonically before disappearance at the critical points. Further more, HEPM thermodynamic scalar curvature differences between SBHs and LBHs are found to decrease monotonically to zero when approaching to the critical points, which is similar as a RN-AdS black hole. We propose that both the number density difference and the HEPM scalar curvature difference can be order parameters describing the SBH/LBH phase transition and judging the upcoming of critical point where a second-order phase transition takes place.  These results provide us with new recognition of the massive gravity. The thermodynamics in the extended phase space of AdS black holes is enriched.
\end{abstract}


\maketitle


\section{Introduction}
Hawking and Page discovered the so-called Hawking-Page phase transition of  a Schwarzschild AdS black hole in 1980s \cite{ref-12}. Since then, phase transitions of AdS black holes have been an important topic in the study of black hole physics. In 1999, Chamblin et al. proposed the first-order phase transition of a RN-AdS black hole, which exhibits the seemingly analogy to a liquid-gas phase transition of classical Van der Waals fluid \cite{ref-13,ref-14}. In 2012, Kubiznak et al. considered the cosmological constant as pressure and its dual quantity as thermodynamic volume, showing us more definitely the analogy of phase transitions between AdS black holes and liquid-gas system \cite{ref-15}. Following their pioneering ideas, lots of meaningful works have been done on the investigation of AdS black holes \cite{Hendi:2012um,Cai:2013qga,Altamirano:2013uqa,Mo:2014wca,Johnson:2014yja,Lan:2015bia,Hendi:2015cra,Fernando:2016qhq}.

Recently, S. Wei et al. projected an insight into the microscopic structure of a RN-AdS black hole based on its thermodynamical phase transition in the extended phase space \cite{ref-20}. They proposed two important quantities, i.e., number density $n$ and specific volume $v$, to describe the microscopic degrees of freedom of the black hole in the phase transition. They found that the number density suffers a sudden change accompanied by a latent heat when the black hole system crosses the SBH/LBH coexistent curve. When the system passes the critical point, it encounters a second-order phase transition with a vanishing latent heat due to the continuous change of the number density. J. Mo et al. extended their work from the RN-AdS black hole to $f(R)$ black holes and Gauss-Bonnet AdS black holes
in modified gravitational theories \cite{ref-21}. It is found that both the coexistent curve and the difference of the number densities of a charged $f(R)$ AdS black hole coincide with those of a RN-AdS black hole while an uncharged  Gauss-Bonnet AdS black hole behaves differently, which may be attributed to the non-trivial effects of the electric charge.

A graviton is massless in the framework of the classical gravitational theory,  numerous facts have supported the validity of this theory. However, due to  hierarchy problem, cosmological problem, accelerated inflation of the universe, etc., scientists are searching for the effective alternatives to general relativity all the time. One of these modified gravitational theory is massive gravity. The basic idea of the massive gravity theory is that gravity is propagated by a spin-2  massive particle \cite{ref-03}. As we know, the introduction of the cosmological constant $\Lambda$ is aimed to explain the acceleration of the universe. However, as the existence of various larger scales from particle physics, such efforts seem helpless. Then with a given value of $\Lambda$, the introduction of a parameter $m$ which has the dimensionality of mass may be helpful, since this particle-physics-independent parameter can explain the accelerated inflation of the universe \cite{ref-04} more reasonably.

Recently, many exact solutions of massive gravity have been proposed \cite{ref-05,ref-06,ref-07,ref-08,ref-09,ref-10}, which all arise at decoupling limits $M_{f}\rightarrow \infty$. The existence of black holes tells us some accurate information of the gravitational theory. The phenomena and properties of black holes are more and more crucial, as they play key roles for our considerations of their corresponding gravitational theories.  There are several works on phase transitions of the massive black holes. Ping Li et al. provided us with static spherically symmetric solutions which may be candidates of black holes in dRGT massive gravity \cite{ref-02}. R. Cai et al. presented a class of charged black hole solutions in an (n+2)-dimensional  holographic massive gravity, and studied the thermodynamics and phase structure of the black hole solutions in both the grand canonical and canonical ensembles \cite{ref-01}. J. Xu et al. further studied the P-V criticality and phase transitions of the holographic massive black holes in canonical ensemble \cite{ref-23}.

In this paper, we would like to study the coexistent physics of dRGT massive black holes and holographic massive black holes in Van der Waals like SBH/LBH phase transitions in the extended phase space where the cosmological constant is viewed as pressure. In Sec.\ref{section two}, we will study the coexistent physics of dRGT black holes along the coexistent curve. In Sec.\ref{section one}, we will investigate the coexistent physics of holographic massive black holes along the coexistent curve. Sec.\ref{section three} will be dedicated to conclusions and discussions.

\section{Coexistent physics of \lowercase{d}RGT massive black holes}\label{section two}
The action of dRGT massive gravity is
\begin{equation}
S=\int d^{4}x\frac{\sqrt{-g}}{16\pi}\left[R-\frac{F^{2}}{4}+m^{2}U(g,\phi^{a})\right].
\end{equation}
Here the potential $U(g,\phi^{a})$ for the graviton as the modification of the gravitational sector can be read as
\begin{equation}
U(g,\phi^{a})=U_{2}+\alpha_{3}U_{3}+\alpha_{4}U_{4},
\end{equation}
$\alpha_{3}$ and $\alpha_{4}$ are the dimensionless parameters, and
\begin{equation}
U_{2}=[K]^{2}-[K^{2}],\\ \nonumber
\end{equation}
\begin{equation}
U_{3}=[K]^{3}-3[K][K^{2}]+2[K^{3}],\\ \nonumber
\end{equation}
\begin{equation}
U_{4}=[K]^{4}-6[K]^{2}[K^{2}]+8[K][K^{3}]+3[K^{2}]^{2}-6[K^{4}],\nonumber
\end{equation}
where $[K]=K^{\mu}{}_{\mu}$
and
\begin{equation}
K^{\mu}{}_{\nu}=\delta^{\mu}{}_{\nu}-\sqrt{g^{\mu\alpha}\partial_{\alpha}\phi^{a}\partial_{\nu}\phi^{b}\eta_{ab}}= \delta^{\mu}{}_{\nu}-\sqrt{\Sigma}^{\mu}{}_{\nu},
\end{equation}
where ${\phi}^{a}$ is the Stückelberg scalar. It is verified that there are several static spherically symmetric solutions \cite{ref-02}
\begin{equation}
ds^{2}=-b^{2}(r)dt^{2}+\frac{1}{a^{2}(r)}dr^{2}+r^{2}d\Omega^{2},
\end{equation}
\begin{equation}
\phi^{0}=t+h(r),
\end{equation}
\begin{equation}
\phi^{i}=\theta x^{i},
\end{equation}
which includes the well-known Reissner-Nordström black hole. In case of
\begin{equation}\label{dRGTalphalimitation}
\theta=\frac{2}{3\alpha_{3}}+1,~~\alpha_{4}\neq \frac{5}{16}{\alpha_{3}}^{2}-\frac{\alpha_{3}}{6},~~\alpha_{4}\neq \frac{3}{8}{\alpha_{3}}^{2}-\frac{\alpha_{3}}{8},
\end{equation}
and with the cosmological constant considered, the static spherically symmetric solution with electric charge of dRGT massive gravity is \cite{ref-02}
\begin{equation}
ds^{2}=-f(r)dt^{2}+\frac{dr^{2}}{f(r)}+r^{2}d\Omega^{2},
\end{equation}
where
\begin{figure*}
   \centering
  \includegraphics[angle=0,width=0.98\textwidth]{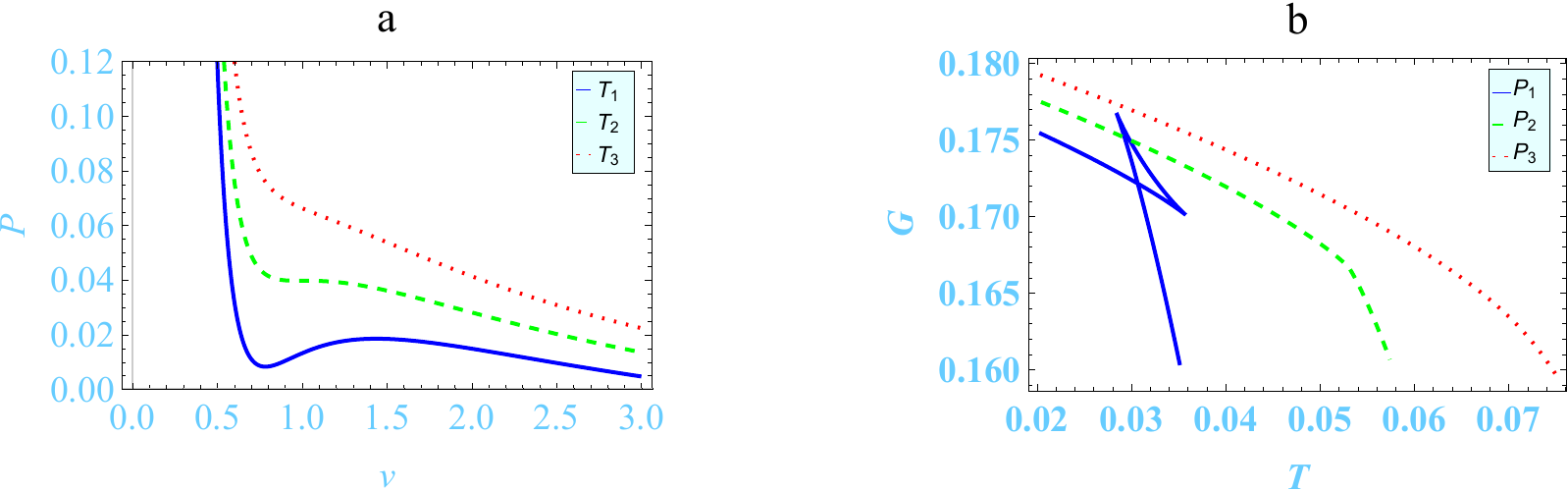}
   \caption{$P-v$ diagram (a) and $G-T$ diagram (b) of dRGT massive black holes for $\beta =1,~~Q=\frac{\sqrt{2}}{12},~~m=\frac{1}{2}$, and $T_{1}=\frac{1}{12\pi},T_{2}=\frac{1}{6\pi},T_{3}=\frac{3}{12\pi}$ (a), $P_{1}=\frac{1}{16\pi},P_{2}=\frac{1}{8\pi},P_{3}=\frac{3}{16\pi}$ (b).}
  \label{figone}
\end{figure*}
\begin{equation}\label{originalmetricfunction}
f(r)=1-\frac{\mu}{r}+\frac{\nu }{r^2}-\beta r^{-\lambda }+\frac{4 m^2 r^2}{27 \alpha _3^2}+\frac{3\alpha _3 r^2}{l^{2}},
\end{equation}
\begin{equation}
\lambda=2 \left(\frac{6 \alpha _3^2+4\alpha _3}{9 \alpha _3^2-16\alpha _4}-1\right), ~\nu =\frac{q^2(3 \alpha _3^2+2\alpha _3)}{2 \left(3 \alpha _3^2-\alpha _3-8\alpha _4\right)}.\nonumber
\end{equation}
Here $\mu,~\beta,~q,~m,~l$ stand for integral constant related to mass of the black hole, scalar charge related to the graviton, electric charge, mass of the graviton, and the curvature radius of AdS space, respectively.

It can be argued that there may exist horizons with proper parameters. The outer horizon can be obtained from $f(r_{+})=0$, where $r_{+}$ stands for the outer horizon radius. According to the Hamiltonian approach, one can obtain the ADM mass $M$ and charge of $Q$ of the black hole as
\begin{equation}
M=\frac{\mu}{2},
\end{equation}
\begin{equation}
Q=\frac{\left(2 \alpha _3+3 \alpha _3^2\right)q}{2 r \left(-\alpha _3+3 \alpha _3^2-8 \alpha _4\right)}.
\end{equation}
After setting $l^{2}=-\frac{3}{\Lambda}=\frac{3}{8 \pi P}$,
the mass, Hawking temperature, entropy,  electric potential, pressure, and thermodynamic volume of the black hole can be obtained routinely as
\begin{equation}
M=\frac{r_{+}}{2}-\frac{r_{+}^{1-\lambda } \beta }{2 (1-\lambda )}+\frac{r_{+}^{1-\lambda } \beta  \lambda }{2 (1-\lambda )}+\frac{\nu }{2 r_{+}}+\frac{2
m^{2} r_{+}^3}{27 \alpha _3^2}-4 P \pi  r_{+}^3 \alpha _3,
\end{equation}
\begin{equation}
T=\frac{1}{4 \pi  r_{+}}-\frac{r_{+}^{-1-\lambda } \beta }{4 \pi }+\frac{r_{+}^{-1-\lambda } \beta  \lambda }{4 \pi }-\frac{\nu }{4 \pi  r_{+}^3}+\frac{m^{2}
r_{+}}{9 \pi  \alpha _3^2}-6 P r_{+} \alpha _3,
\end{equation}
\begin{equation}
S=\pi r_{+}^{2},
\end{equation}
\begin{equation}
\Phi=\frac{2 \left(2 \alpha _3+3 \alpha _3^2\right) \left(-\alpha _3+3 \alpha _3^2-8 \alpha _4\right) Q}{\alpha _3^2 \left(2+3 \alpha
_3\right){}^2 r_{+}},
\end{equation}
 \begin{equation}
P=\frac{m^{2}}{54 \pi  \alpha _3^3}+\frac{1}{24 \pi  r_{+}^2 \alpha _3}-\frac{T}{6 r_{+} \alpha _3}+\frac{r_{+}^{-2-\lambda } \beta}{24 \pi  \alpha _3}(\lambda-1)-\frac{\nu }{24 \pi  r_{+}^4 \alpha _3},
\end{equation}
\begin{equation}\label{volumedRGT}
V=-4 \pi  r_{+}^3 \alpha _3.
\end{equation}

Besides, we would like to introduce two extra quantities $V_{m}$, $\Phi_{\beta}$ conjugated to $m^{2}$ and $\beta$ as
\begin{equation}
V_{m}=\frac{2 r_{+}^3}{27 \alpha _3^2},
\end{equation}
\begin{equation}
\Phi_{\beta}=\frac{r_{+}^{1-\lambda } (-1+\lambda )}{2 (1-\lambda )}.
\end{equation}
Different values of $\lambda$ lead to different spacetime geometries. The corresponding Ricci scalar of the metric is
\begin{equation}
R=-\frac{12}{l^2}-16 m^2+\beta (\lambda^{2}-3\lambda+2) r^{-2-\lambda }.
\end{equation}
Therefore, we can know that when $\lambda\geqslant-2$, the asymptotical behaviour of this solution at $r\rightarrow \infty$ is AdS. Now we will consider the case $\lambda=-1$, which is sufficiently representative for exhibiting the coexistent thermodynamic properties of dRGT massive black holes. After setting the value of the parameter $\alpha_{3}=-1/3$ (such that the thermodynamic volume in the following Eq. (\ref{volumedRGT}) can be $4 \pi  r_{+}^3/3$ which coincides with that of general AdS black holes), the metric function Eq. (\ref{originalmetricfunction}) can then be rewritten as
\begin{equation}
f(r)=1-\frac{2 M}{r}+\frac{3Q^2}{r^2}-\beta r+\frac{4 m^2 r^2}{3}+\frac{8}{3}\pi P  r^2.
\end{equation}
At the same time, we have noticed that $\alpha_{4}=7/48$ is needed, which satisfies the limitation Eq. (\ref{dRGTalphalimitation}).

\begin{figure*}
   \centering
  \includegraphics[angle=0,width=0.98\textwidth]{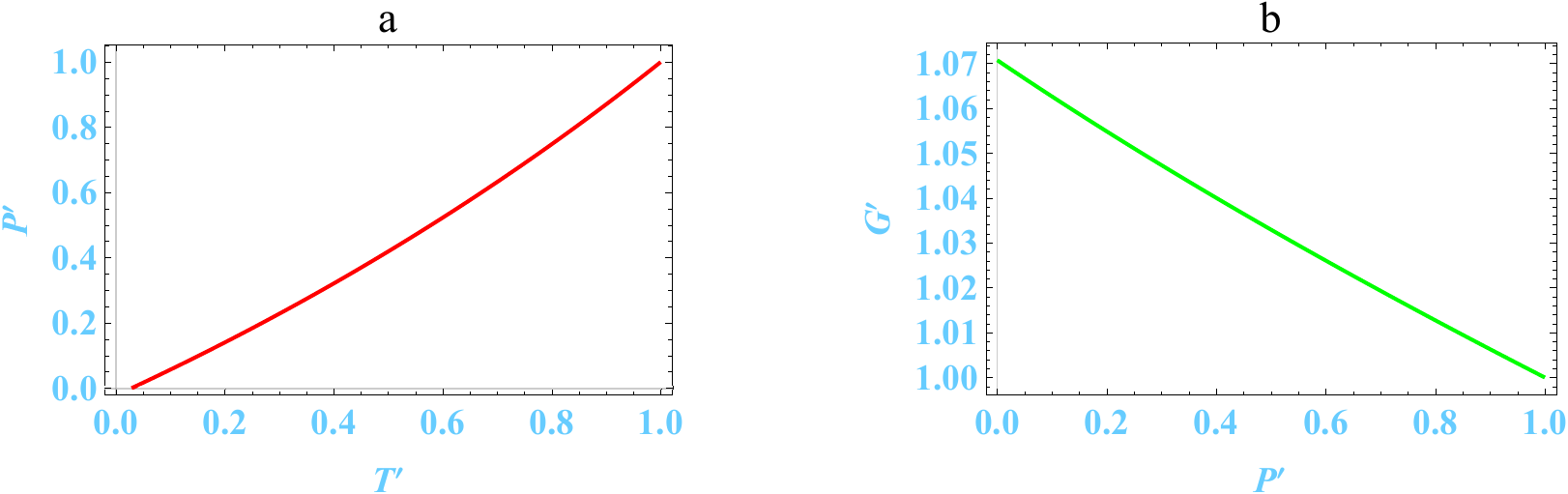}
   \caption{$P^{\prime}-T^{\prime}$ coexistent diagram (a) and $G^{\prime}-P^{\prime}$ coexistent diagram (b) of dRGT massive black holes for $\beta =1,~~Q=\frac{\sqrt{2}}{12},~~m=\frac{1}{2}$.}
  \label{figtwo}
  \end{figure*}

Therefore, the first law of the black hole is then conformed as
\begin{equation}
dM=TdS+VdP+\Phi dQ+V_{m}dm^{2}+\Phi_{\beta} d\beta,
\end{equation}
and the corresponding Smarr relation reads
\begin{equation}
M=2TS-2VP+\Phi Q-2V_{m}m^{2}-\Phi_{m} \beta.
\end{equation}

What follows up is the phase transition of the black holes.
Introducing the definition of the specific volume as
\begin{equation}\label{defofspecificvolume}
v=2r_{+},
\end{equation}
the equation of state can be re-expressed as
\begin{equation}
P=\frac{36 Q^2+3 v^2 \left(-1+2 \pi  T v-m^2 v^2+v \beta \right)}{6 \pi  v^4}.
\end{equation}
Considering the canonical ensemble with a fixed electric charge $Q$, the Gibbs free energy reads
\begin{equation}
G=M-TS=\frac{108 Q^2+3 v^2-m^2 v^4-2 P \pi  v^4}{24 v}.
\end{equation}

The corresponding $P-v$ diagram and $G-T$ diagram have been shown in Fig.\ref{figone}. The oscillatory behaviour in the $P-v$ diagram and the swallow tail phenomenon in the $G-T$ diagram show that there is a Van der Waals like first-order phase transtion for $T<T_{c}$ or $P<P_{c}$, with $T_{c}$ and $P_{c}$ standing for the critical value of the phase trasition respectively.  The analytical critical point of the black hole in the extended phase space can be obtained from
 \begin{equation}\label{critical}
\frac{\partial P}{\partial v}=0,~~\frac{\partial^{2} P}{\partial v^{2}}=0,
\end{equation}
which leads to
\begin{equation}
v_{c}=6\sqrt{2}Q,~~T_{c}=\frac{\sqrt{2}-9 Q \beta }{18 \pi  Q},\nonumber
\end{equation}
\begin{equation}
P_{c}=\frac{1-144 m^2 Q^2}{288 \pi  Q^2},~~G_{c}=\sqrt{2}Q,\nonumber
\end{equation}
then the critical value is
\begin{equation}
\rho_{c}=\frac{P_{c}v_{c}}{T_{c}}=\frac{3-432 m^2 Q^2}{16-72 \sqrt{2} Q \beta }.
\end{equation}
 We have set the parameters as $\beta =1,~~Q=\sqrt{2}/12,~~m=1/2$ in Fig. \ref{figone}, so that $\rho_{c}=3/8$, which is the same value as that of the Van der Waals fluid in liquid/gas phase transitions.

 \begin{figure*}
      \includegraphics[angle=0,width=0.98\textwidth]{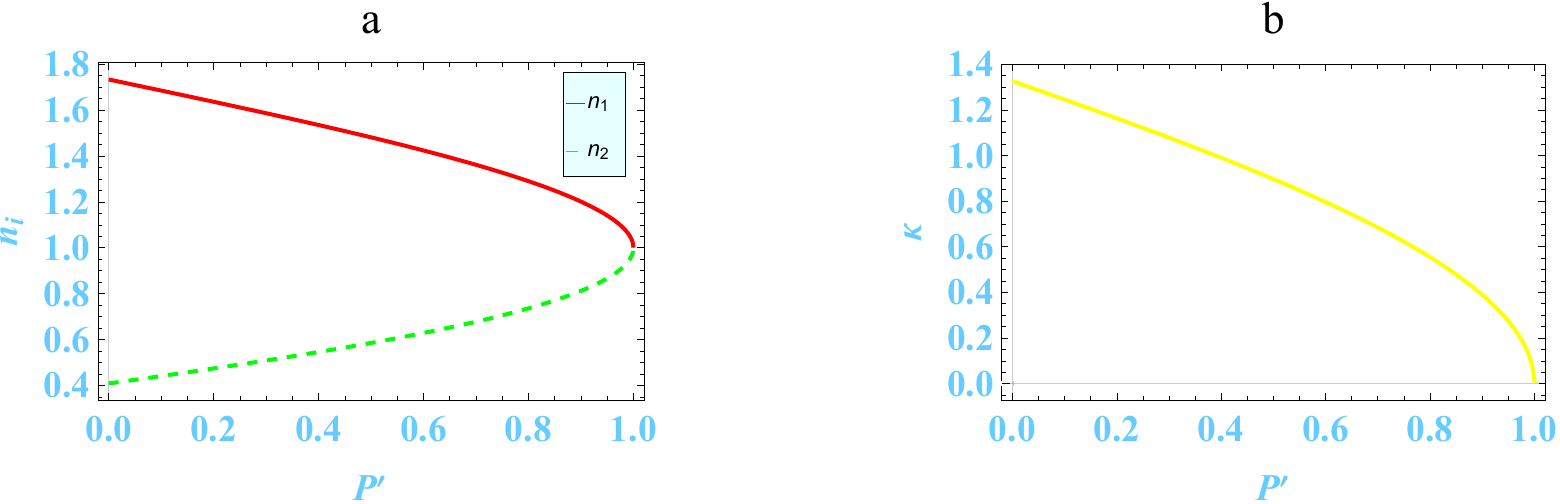}
   \caption{Number density of two phases of dRGT black hole along the coexistent curve (a) and number density difference between the two phases of dRGT black hole (b) for $\beta =1,~~Q=\frac{\sqrt{2}}{12},~~m=\frac{1}{2}$.}
  \label{figthree}
\end{figure*}

After defining the reduced thermodynamic quantities as
\begin{equation}\label{reduced def}
v^{\prime}\equiv\frac{v}{v_{c}},~~T^{\prime}\equiv\frac{T}{T_{c}},~~ P^{\prime}\equiv\frac{P}{P_{c}},~~ G^{\prime}\equiv\frac{G}{G_{c}},
\end{equation}
we can then obtain the reduced temperature and reduced  Gibbs free energy as
\begin{equation}
T^{\prime}(v^{\prime})=\frac{-1+6 {v^{\prime}}^2-3Z {v^{\prime}}^4-36 \sqrt{2} Q {v^{\prime}}^3 \beta }{4
{v^{\prime}}^3 \left(2-9 \sqrt{2} Q \beta \right)},
\end{equation}
\begin{equation}
G^{\prime}(v^{\prime})=\frac{3+6{v^{\prime}}^2+Z {v^{\prime}}^4}{8 v^{\prime}},
\end{equation}
where $Z=P^{\prime}(-1+144 m^2 Q^2)-144m^{2}Q^{2}$.

When the temperature is less than the critical value, the phase transition will happen, during which there will be the coexistence of SBHs and LBHs. We now want to obtain the coexistent curve of  dRGT massive black hole. It is supposed that there are two phases at any point on the coexistent curve, with their own thermodynamic quantities  $v_{1}^{\prime},~P_{1}^{\prime},~T_{1}^{\prime},~G_{1}^{\prime}$ and $v_{2}^{\prime},~P_{2}^{\prime},~T_{2}^{\prime},~G_{2}^{\prime}$, respectively. Due to the continuity of Gibbs free energy and temperature from one phase to the other one,  we have
\begin{equation}\label{eq2}
T^{\prime}(v_{1})=T^{\prime}(v_{2})=T^{\prime},
\end{equation}
\begin{equation}\label{eq1}
G^{\prime}(v_{1})=G^{\prime}(v_{2})=G^{\prime},
\end{equation}
\begin{equation}\label{eq4}
T^{\prime}(v_{1})+T^{\prime}(v_{2})=2T^{\prime},
\end{equation}
\begin{equation}\label{eq3}
G^{\prime}(v_{1})+G^{\prime}(v_{2})=2G^{\prime}.
\end{equation}

With Eqs. (\ref{eq2}), (\ref{eq1}) and (\ref{eq4}) , we can analytically obtain the equation of the coexistent curve, which reads
\begin{equation}
T^{\prime}=\frac{\left[ \left(1-144 m^2 Q^2\right) AP^{\prime}+ 3\left(144 m^2 Q A-3 \beta \right)Q\right]}{\sqrt{2}-9 Q \beta },
\end{equation}
where
\begin{equation}
A=\sqrt{\frac{3-\sqrt{P^{\prime}+144 m^2 Q^2-144 m^2 P^{\prime} Q^2}}{P^{\prime}+144
m^2 Q^2-144 m^2 P^{\prime} Q^2}}.\nonumber
\end{equation}
With Eqs. (\ref{eq2}), (\ref{eq1}) and (\ref{eq3}), we  can obtain the reduced Gibbs free energy along the coexistent curve as
\begin{equation}
G^{\prime}=A\sqrt{72 m^2 Q^2+P^{\prime} \left(\frac{1}{2}-72 m^2 Q^2\right)}.
 \end{equation}

The $P^{\prime}-T^{\prime}$ diagram and $G^{\prime}-P^{\prime}$ diagram are shown in Fig.\ref{figtwo}. From the $P^{\prime}-T^{\prime}$ diagram, we can see that the curve does not pass through the point (0, 0), which is different from that of RN-AdS black holes obtained in Ref. \cite{ref-24}  and $f(R)$ black holes obtained in Ref. \cite{ref-21}. It is obvious that there will be no coexistence of SBH and LBH when the temperature is close to zero. On the other hand, the reduced Gibbs free energy declines monotonically with the increasing reduced pressure $P^{\prime}$.

\begin{figure*}
   \centering
    \includegraphics[angle=0,width=0.98\textwidth]{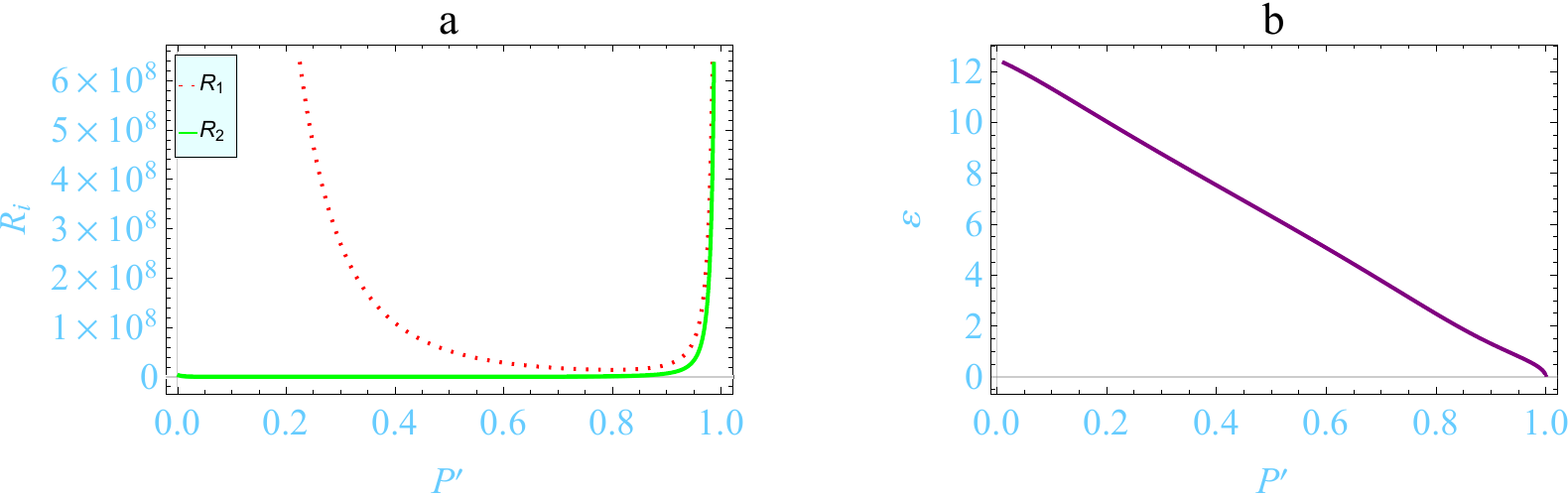}
   \caption{HEPM thermodynamical scalar curvatures of two phases of dRGT black hole along the coexistent curve (a)  and the HEPM scalar curvature ratio of the two phases of dRGT black hole (b) for $\beta =1,~~Q=\frac{\sqrt{2}}{12},~~m=\frac{1}{2}$.}
  \label{figfour}
\end{figure*}

Referring to Ref. \cite{ref-20}, we can define a quantity
\begin{equation}\label{molecule def}
n\equiv\frac{1}{v},
\end{equation}
which represents number density of a black hole. Then using Eqs. (\ref{eq2}) and (\ref{eq1}),  we can obtain the number density of the SBHs and LBHs as
\begin{equation}\label{density1}
n_{1}=\left(\frac{6}{1+P'}-\frac{2 \sqrt{2}}{\sqrt{1+P^{\prime}}}-\frac{1}{2} \sqrt{B}\right)^{-\frac{1}{2}},
\end{equation}
\begin{equation}\label{density2}
n_{2}=\left(\frac{6}{1+P'}-\frac{2 \sqrt{2}}{\sqrt{1+P^{\prime}}}+\frac{1}{2} \sqrt{B}\right)^{-\frac{1}{2}},
\end{equation}
where
\begin{equation}
B=-\frac{4}{1+P^{\prime}}-\frac{86\sqrt{2}}{(1+P^{\prime})^{3/2}}+ \frac{172+28 P^{\prime}}{(1+P')^2}.\nonumber
\end{equation}
One should notice that we have set $\beta =1,~~Q=\frac{\sqrt{2}}{12},~~m=\frac{1}{2}$ here for simplicity.

It is shown in Fig. \ref{figthree} that with the increasing pressure the  number density of SBH decreases while the other one increases. When the critical point is arrived at, the number densities of SBHs and LBHs become equal.

To further reveal the number density behaviour along the coexistent curve, we can define a quantity \cite{ref-21}
\begin{equation}\label{molecule difference}
\kappa\equiv n_{1}-n_{2},
\end{equation}
which reflects the difference of the number density between the two black holes in the phase transition. Considering Eqs. (\ref{density1}) and (\ref{density2}), we can obtain
\begin{equation}
\kappa=\sqrt{6-6\sqrt{P^{\prime}+144 m^2 Q^2-144 m^2 Q^2 P^{\prime}}}.
\end{equation}
We have shown the relation between $P^{\prime}$ and $\kappa$ in Fig. \ref{figthree}.
It tells us explicitly that the number density difference between SBH and LBH becomes smaller and smaller with the increasing $P^{\prime}$, and the difference disappears at the critical point. This indicates that  $\kappa$ is an order parameter of the phase transition.

Now we would like to investigate the thermodynamic scalar curvature of dRGT massive black holes in the Van der Waals like phase transitions. The recently proposed HEPM thermodynamical metric is defined as \cite{Hendi:2015rja}
\begin{equation}
ds^{2}=S\frac{M_{S}}{M_{QQ}^{3}}(-M_{SS}dS^{2}+M_{QQ}dQ^{2}).
\end{equation}
Accordingly, we can obtain the scalar curvature
\begin{equation}\label{scalarcurvatureone}
R=\frac{X(3)}{X(1)X(2)},
\end{equation}
where
\begin{equation}
X(1)=2187 Q^5 \left(-1+2 {v^{\prime}}^2+Z{v^{\prime}}^4\right)^2,\nonumber
\end{equation}
\begin{equation}
X(2)=\left(1+3Z {v^{\prime}}^4+6 {v^{\prime}}^2 \left(-1+3 \sqrt{2} Q {v^{\prime}} \beta \right)\right)^3.\nonumber
\end{equation}

Applying Eqs. (\ref{molecule def}), (\ref{density1}), (\ref{density2}) and (\ref{scalarcurvatureone}), scalar curvatures of SBH and LBH can be derived, which are intuitively shown in Fig.\ref{figfour}, where $R_{1}$, $R_{2}$ stand for scalar curvature of SBH and LBH respectively. With the increasing $P^{\prime}$, one can see that the curvature of LBH increases slowly while that of SBH decreases drastically. However, both of them increase to infinity when $P^{\prime}$ is approaching to the critical value. One may naturally compare these two curvatures for more profound information, which leads to a new quantity, i.e. the scalar curvature ratio $\varepsilon$ between SBH and LBH defined as
\begin{equation}
\varepsilon=\ln\frac{R_{1}}{R_{2}}.
\end{equation}
After visualization as Fig.\ref{figfour}, we can find that the scalar curvature ratio $\varepsilon$ increases mononically with the increasing $P^{\prime}$, which means that the closer to the critical point, the smaller difference of the scalar curvatures between the two phases. We can also find that the scalar curvature difference disappears just at the critical point, which tells us that the two phases become unrecognisable and identical.

\begin{figure*}
  \includegraphics[angle=0,width=0.98\textwidth]{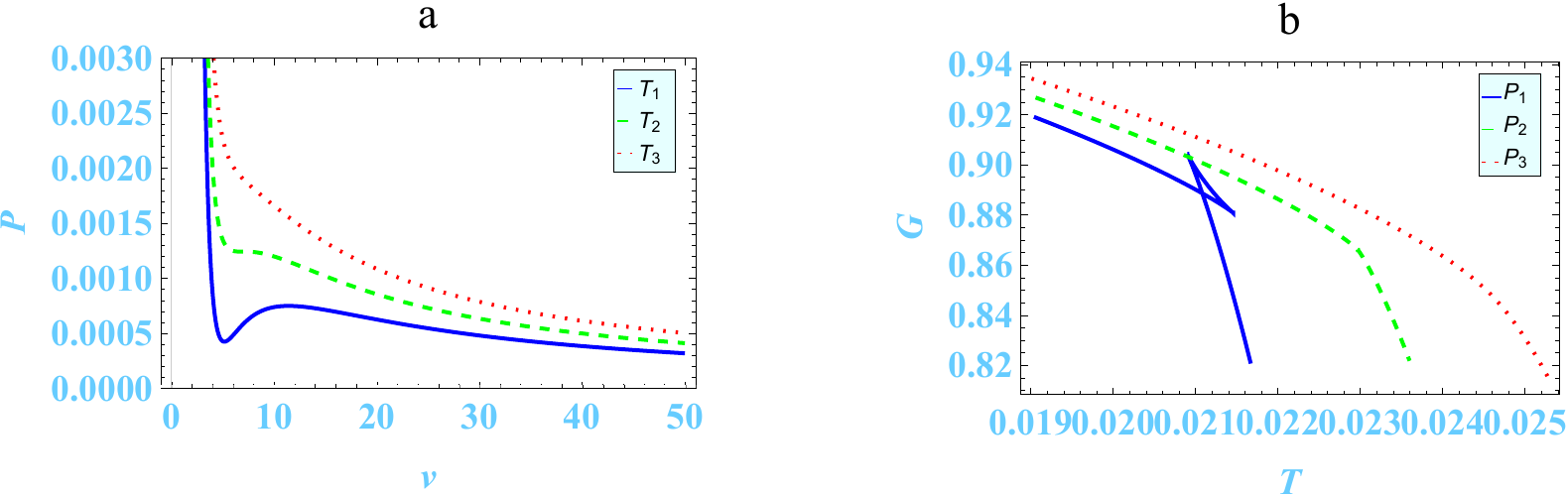}
   \caption{$P-v$ diagram (a) and $G-T$ diagram (b) of the holographic massive black holes for $c_1=1, c_2=-1, Q=1, m=\frac{1}{2}$ and $T_{1}=\frac{4}{40\sqrt{3}\pi},T_{2}=\frac{5}{40\sqrt{3}\pi},T_{3}=\frac{6}{40\sqrt{3}\pi}$ (a), $P_{1}=\frac{1}{320\pi},P_{2}=\frac{1}{256\pi},P_{3}=\frac{3}{640\pi}$ (b). }
  \label{figfive}
\end{figure*}

\section{Coexistent physics of holographic massive black holes}\label{section one}
The action and static spherically symmetric solutions of $(n+2)$-dimensional massive gravity have been reviewed in Ref. \cite{ref-01} as
\begin{equation}
S=\int d^{n+2}x\frac{\sqrt{-g}}{16\pi}\left[R+\frac{n(n+1)}{l^{2}}-\frac{F^{2}}{4}+m^{2}U(g, \phi^{a})\right],
\end{equation}
\begin{equation}
ds^{2}=-f(r)dt^{2}+\frac{1}{f(r)}dr^{2}+r^{2}h_{ij}dx^{i}dx^{j},
\end{equation}
where
\begin{equation}\label{holographic metric}
\begin{aligned}
f(r)&=k+\frac{16 \pi  P}{(n+1) n} r^2-\frac{16 \pi  M}{n \Omega r^{n-1}}\\&+\frac{(16 \pi  Q)^2}{2 n (n-1) \Omega^2 r^{2(n-1)}}+\frac{c_0 c_1m^2}{n}
r+c_0^2 c_2 m^2\\&+
\frac{(n-1) c_0^3 c_3 m^2}{r}+\frac{(n-1) (n-2) c_0^4 c_4 m^2}{r^2}.
\end{aligned}
\end{equation}
Here $c_{i}(i=1, 2, 3, 4),~\Omega,~M,~Q$ are the dimensionless parameters of the potential $U(g, \phi^{a})$, the volume of space spanned by the coordinates $x^{i}$, the mass of the black hole and the charge of the black hole, respectively. We have viewed the cosmological constant as a variable, identifying it with the pressure as $P=n(n+1)/16\pi l^{2}$.
This solution is based on the reference metric $f_{\mu\nu}=diag(0,0,c_{0}^{2}h_{ij}),$
where $c_{0}$ is a positive constant which, without loss of generality, can be set as $c_{0}=1$ \cite{ref-01}. One may take $k=-1,0$ or $ 1$, relating to a spherical, Ricci flat, or hyperbolic horizon of the black hole. We would like to consider the case of  4-dimensional spherically symmetric black holes, which may lead to $U_{3}=U_{4}=0$. As a result, we can naturally set $c_{3}=c_{4}=0$ \cite{ref-23} and the metric function can be reduced to
\begin{equation}
f(r)=1-\frac{2 M}{r}+\frac{4 Q^2}{r^2}+\frac{8}{3} P \pi  r^2+\frac{1}{2} m^2 c_1 r+m^2 c_2.
\end{equation}

The black hole horizon is determined by the constraint $f(r_{+})=0$. Then the mass, Hawking temperature, entropy, electric potential, pressure, thermodynamic volume and Gibbs free energy in the canonical ensemble of the black hole can be given as \cite{ref-23}
\begin{equation}
M=\frac{24 Q^2+6 r_{+}^2+16 P \pi  r_{+}^4+3 m^2 r_{+}^3 c_1+6 m^2 r_{+}^2  c_2}{12 r_{+}},
\end{equation}
\begin{equation}
T=\frac{-4 Q^2+r_{+}^2+8 P \pi  r_{+}^4+m^2 r_{+}^3 c_1+m^2 r_{+}^2  c_2}{4 \pi  r_{+}^3},
\end{equation}
\begin{equation}
S=\pi r_{+}^{2},
\end{equation}
\begin{equation}
\Phi=\frac{4Q}{r_{+}},
\end{equation}
\begin{equation}
P=\frac{4 Q^2-r_{+}^2+4 \pi  r_{+}^3 T-m^2 r_{+}^3 c_1-m^2 r_{+}^2  c_2}{8 \pi  r_{+}^4},
\end{equation}
\begin{equation}\label{volume}
V=\frac{4 \pi  r_{+}^3}{3},
\end{equation}
\begin{equation}
G=\frac{3 Q^2}{r_{+}}+\frac{r_{+}}{4}-\frac{2}{3} P \pi  r_{+}^3+\frac{1}{4} c_2 m^2 r_{+}.
\end{equation}

\begin{figure*}
  \includegraphics[angle=0,width=0.98\textwidth]{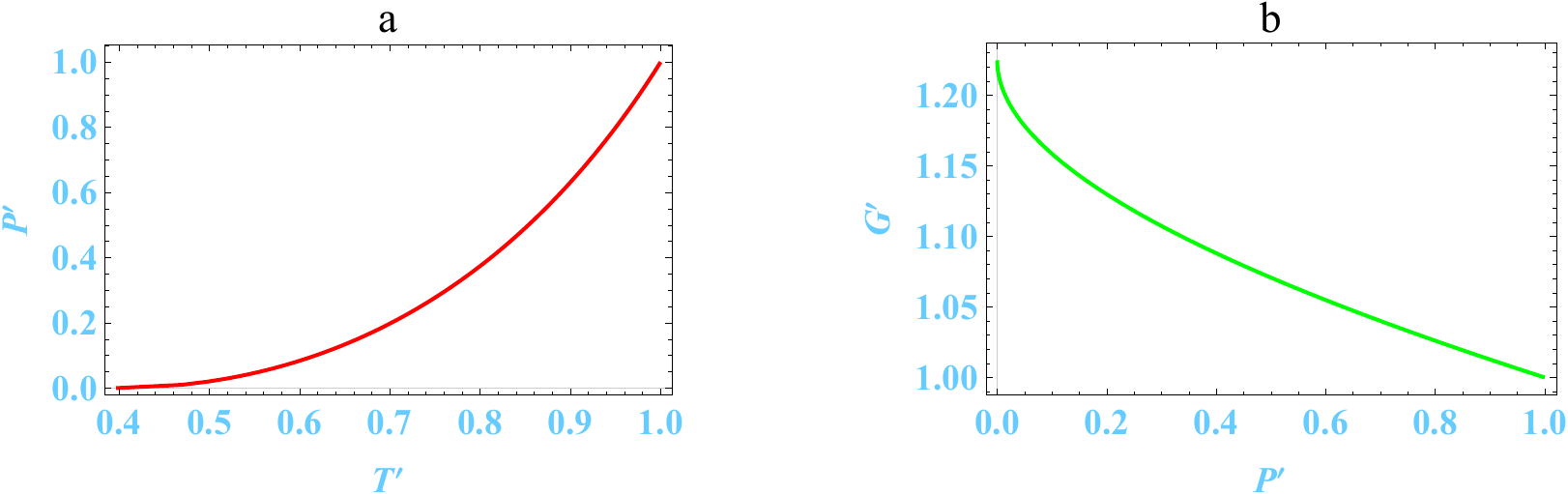}
   \caption{$P^{\prime}-T^{\prime}$ coexistent diagram (a)  and $G^{\prime}-P^{\prime}$ coexistent diagram (b) of holographic massive black hole for $ c_{1}=1, ~c_2=-1,~Q=1,~m=\frac{1}{2}$.}
\label{figsix}
\end{figure*}

One can check that the thermodynamic quantities obey the following differential formula \cite{ref-23}
\begin{equation}
dM=TdS+VdP+\Phi dQ+\frac{m^{2}r^{2}}{4}dc_{1}+\frac{m^{2}r}{2}dc_{2},
\end{equation}
where $c_{1}, c_{2}$ are viewed as variables so that it coincides with Smarr formula
\begin{equation}
M=2TS-2VP+\Phi Q-\frac{c_{1}m^{2}r^{2}}{4}.
\end{equation}

With the definition of the specific volume by Eq. (\ref{defofspecificvolume}), the equation of state of the black hole can be expressed as
\begin{equation}
P=-\frac{m^2 v^3 c_1+2 \left(-16 Q^2+v^2-2 \pi  T v^3+m^2 v^2  c_2\right)}{4 \pi  v^4}.
\end{equation}
Then the critical point can be obtained as
\begin{equation}
v_{c}=\frac{4 \sqrt{6} Q}{C},\nonumber
\end{equation}
\begin{equation}
T_{c}=\frac{\sqrt{6}+2 \sqrt{6} m^2  c_2+\sqrt{6} m^4 c_2^2+9 m^2 Q c_1 C}{36 \pi  Q C},\nonumber
\end{equation}
\begin{equation}
P_{c}=\frac{C^2}{384 \pi  Q^2},\nonumber
\end{equation}
\begin{equation}
G_{c}=2 \sqrt{\frac{2}{3}} Q C,
\end{equation}
where $C=\sqrt{1+m^2  c_2}$, and the critical value is
\begin{equation}
\rho_{c}=\frac{P_{c}v_{c}}{T_{c}}=\frac{3 \sqrt{6} C^{4}}{8 \left[\sqrt{6}+\sqrt{6} m^2 c_2(1+C^{2})+9 m^2
QC c_1\right]}.
\end{equation}
It is obvious that when $m=0$, $\rho_{c}$ will be $3/8$, which is the same value as that of the Van der Waals fluid in liquid/gas phase transitions.

The phase diagram has been shown in Fig.\ref{figfive}. We can know that there exists a first-order phase transition of the black hole for $T<T_{c}$ or $P<P_{c}$.

Using the definition Eq. (\ref{reduced def}), one can get
\begin{equation}
T^{\prime}(v^{\prime})=\frac{12 \sqrt{6} m^2 Q {v^{\prime}}^3 c_1+\left(-1+6 {v^{\prime}}^2+3 p^{\prime} {v^{\prime}}^4\right) {C}^{3}}{4
{v^{\prime}}^3 \left(3 \sqrt{6} m^2 Q c_1+2C^{3}\right)},
\end{equation}

\begin{equation}
G^{\prime}(v^{\prime})=\frac{3+6 {v^{\prime}}^2-p^{\prime} {v^{\prime}}^4}{8 {v^{\prime}}}.
\end{equation}

\begin{figure*}
\includegraphics[angle=0,width=0.98\textwidth]{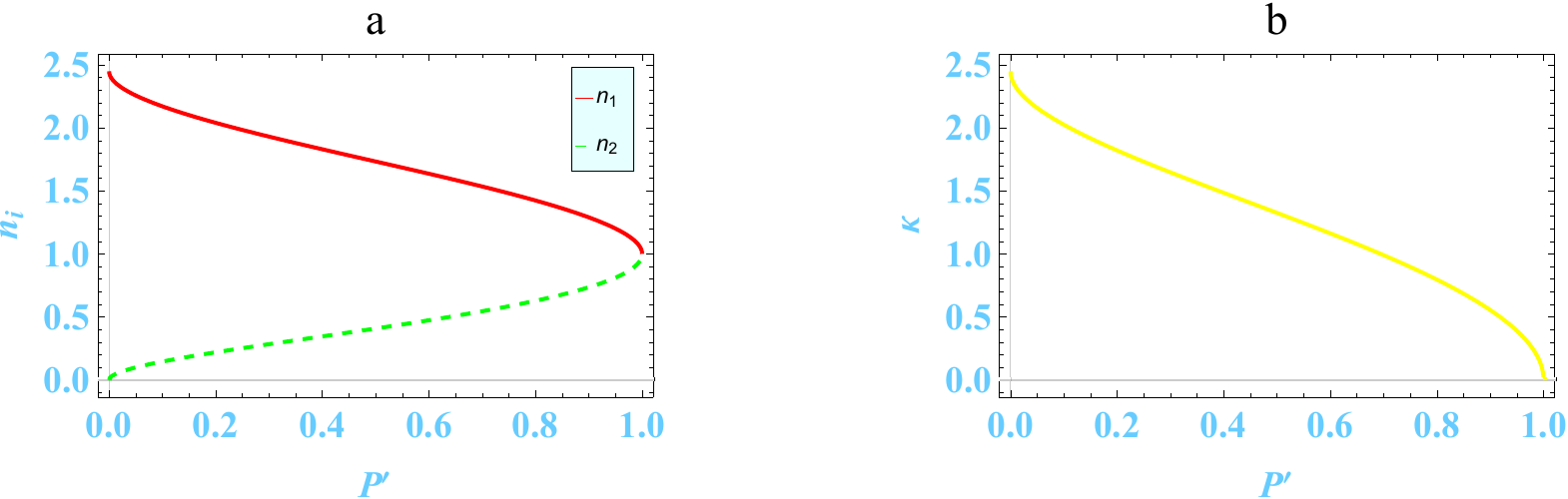}
   \caption{Number density of two phases of holographic massive black hole along the coexistent curve (a) and number density difference of the two phases of holographic massive black hole (b).}
\label{figseven}
\end{figure*}

Since the phase transition is first-order, the Gibbs free energy and the temperature should be equal in the two equilibrial states. Without loss of generality, supposing there are two phases at any point on the coexistent curve, with their own thermodynamic quantities  $v_{1}^{\prime},~P_{1}^{\prime},~T_{1}^{\prime},~G_{1}^{\prime}$ and $v_{2}^{\prime},~P_{2}^{\prime},~T_{2}^{\prime},~G_{2}^{\prime}$ respectively,  we can have
\begin{equation}\label{temp1}
T^{\prime}(v_{1})=T^{\prime}(v_{2})=T^{\prime},
\end{equation}
\begin{equation}\label{Gibbs}
G^{\prime}(v_{1})=G^{\prime}(v_{2})=G^{\prime},
\end{equation}
\begin{equation}\label{Temperature}
T^{\prime}(v_{1})+T^{\prime}(v_{2})=2T^{\prime}.
\end{equation}
\begin{equation}\label{Gibbs2}
G^{\prime}(v_{1})+G^{\prime}(v_{2})=2G^{\prime},
\end{equation}

Using Eqs.(\ref{temp1}), (\ref{Gibbs}) and (\ref{Temperature}), we can obtain the analytical equation fitting the coexistent curve as
\begin{equation}
T^{\prime}=\frac{\sqrt{2} \left(3 \sqrt{3} m^2 Q c_1+\sqrt{\frac{3-\sqrt{P^{\prime}}}{P^{\prime}}} P^{\prime} C^{3}\right)}{3
\sqrt{6} m^2 Q c_1+2 C^{3}}.
\end{equation}
Using Eqs.(\ref{temp1}), (\ref{Gibbs}) and (\ref{Gibbs2}), we can obtain the reduced Gibbs free energy along the coexistent curve as
\begin{equation}
G^{\prime}=\sqrt{\frac{3-\sqrt{P^{\prime}}}{2}}.
\end{equation}
Thinking of the definition Eq. (\ref{molecule def}), we have
\begin{equation}
n_{1}=\left(\frac{3-\sqrt{9-12 \sqrt{P^{\prime}}+3P^{\prime}}}{P^{\prime}}-\frac{2}{\sqrt{P^{\prime}}}\right)^{-\frac{1}{2}},
\end{equation}
\begin{equation}
n_{2}=\left(\frac{3+\sqrt{9-12 \sqrt{P^{\prime}}+3P^{\prime}}}{P^{\prime}}-\frac{2}{\sqrt{P^{\prime}}}\right)^{-\frac{1}{2}}.
\end{equation}
According to the definition of number density difference as Eq. (\ref{molecule difference}), we can obtain
\begin{equation}
\kappa=\sqrt{6-6 \sqrt{P^{\prime}}} .
\end{equation}

The $P^{\prime}-T^{\prime}$ coexistent curve are plotted in Fig.\ref{figsix}. From the diagram we can see intuitively that the coexistent curve does not pass through the point (0,0). It means that when the temperature is small enough, there will be no coexistence of SBH and LBH. The $G^{\prime}-T^{\prime}$ coexistent curve in Fig. \ref{figsix} tells us that the Gibbs free energy in the canonical ensemble decreases monotonically with the increasing pressure, which is the same as that of dRGT massive black holes.

In the microscopic insight, we can see from Fig.\ref{figseven} that the number densities of SBHs and LBHs change in the contrary directions during the phase transitions. The number density difference of the two phases vanishes in the end, when it arrives at the critical point. It conforms that $\kappa$ is the order parameter of the phase transition.

\begin{figure*}
   \centering
    \includegraphics[angle=0,width=0.98\textwidth]{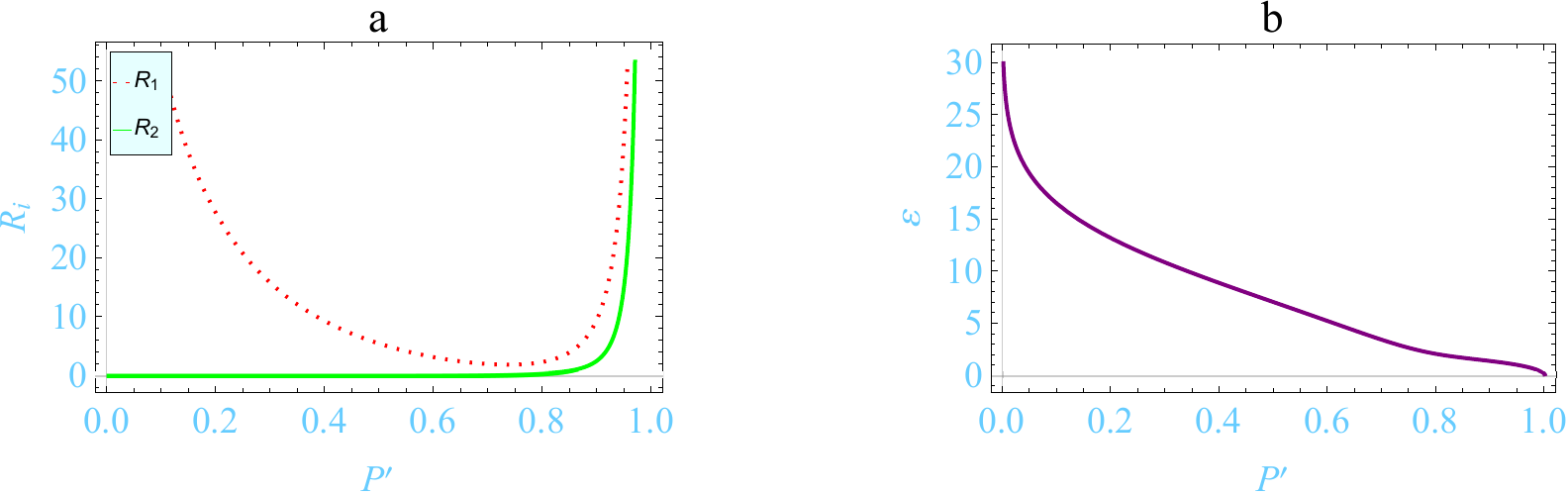}
   \caption{HEPM thermodynamical scalar curvatures of two phases of the holographic black hole along the coexistent curve (a) and the HEPM scalar curvature ratio of the two phases of the black hole (b) for $c_{1} =1,~~c_{2}=-1,~~Q=1,~~m=\frac{1}{2}$.}
  \label{figeight}
\end{figure*}

We can also get the HEPM thermodynamical scalar curvature as
\begin{equation}
R=\frac{Y(3)}{Y(1)Y(2)},
\end{equation}
where
\begin{equation}
Y(1)=3 Q^5 \left(1-2 {v^{\prime}}^2+P^{\prime} {v^{\prime}}^4\right)^2, \nonumber
\end{equation}
\begin{equation}
Y(2)=\left(12 \sqrt{6} {c_1}m^2 Q {v^{\prime}}^3 + C^{3}\left(-1+6 {v^{\prime}}^2+3 P^{\prime} {v^{\prime}}^4\right)\right)^3.\nonumber
\end{equation}
Then the scalar curvature ratio $\varepsilon$ can also be calculated. $R_{1}$, $R_{2}$ and  $\varepsilon$ are shown in Fig.\ref{figeight}. Interestingly, we find that all of them obey the same laws as that of dRGT massive black holes.

\section{Conclusions and discussions}\label{section three}

The coexistent physics of  the dRGT massive black holes and the holographic massive black holes have been investigated in  the extended phase space where the cosmological constant is identified as thermodynamic pressure. In conclusion, we find the following facts:

\romannumeral1. Van der Waals like phase transitions are found in both of them. Due to the non-trivial effect of graviton mass, the $P^{\prime}-T^{\prime}$ coexistent curves, unlike that of RN-AdS black holes, will not pass through (0,0). The $G^{\prime}-P^{\prime}$ coexistent curve shows that Gibbs free energy in the canonical ensemble  decreases monotonically with the increasing pressure, which is similar to general thermodynamic systems.

\romannumeral2. The number densities of the SBHs and LBHs change in the contrary tendencies with the increasing pressure, both of them arrive at the same point when the pressure arrives at the critical point. In other words, the difference of number densities between SBH and LBH becomes smaller and smaller when the pressure approaches to the critical point and this difference becomes zero finally at the critical point, which coincides with that of the RN-AdS black holes \cite{ref-24}, f(R) black holes and Gauss-Bonnet black holes \cite{ref-21}.

\romannumeral3. HEPM thermodynamic scalar curvature of the LBH increases slowly while that of the SBH decreases drastically with increasing $P^{\prime}$. However, both of them increase to infinity when $P^{\prime}$ is approaching to the critical point. After defining
\begin{equation}
\varepsilon\equiv\ln \frac{\text{HEPM scalar curvature of SBH}}{\text{HEPM scalar curvature of LBH}},
\end{equation}
we can directly see that the scalar curvature difference between the two phases, just like the number density difference, will become smaller and smaller as approaching the critical point and finally decrease to zero just at the critical point.

Meanwhile, we want to give some comments as follows.

\romannumeral1. The introduction of the concept of number density, though seemingly naive, may be on some extent reasonable. From the perspective of holographic dual, the microscopic number degrees of freedom $N$ for non-rotating black hole is related to the area of black hole horizon $A$ as  \cite{PhysRevD.78.024016,Altamirano:2014tva}
\begin{equation}
N=\frac{A}{4}\frac{n-2}{n-1}L_{pl}^{-2},
\end{equation}
where $L_{pl}=\sqrt{\hbar G/c^{3}}$ and $n$ is the spacetime dimensions. Then the number density $n$ can be defined as
\begin{equation}
n=\frac{N}{V},
\end{equation}
where $V$ is thermodynamic volume of a non-rotating AdS black hole. So $n$ may be a bridge between microscopic and macroscopic black hole physics quantities in terms of a pure thermodynamic viewpoint. It plays an effective role of discovering interesting phenomenon in phase transitions.

\begin{figure*}
   \centering
    \includegraphics[angle=0,width=0.98\textwidth]{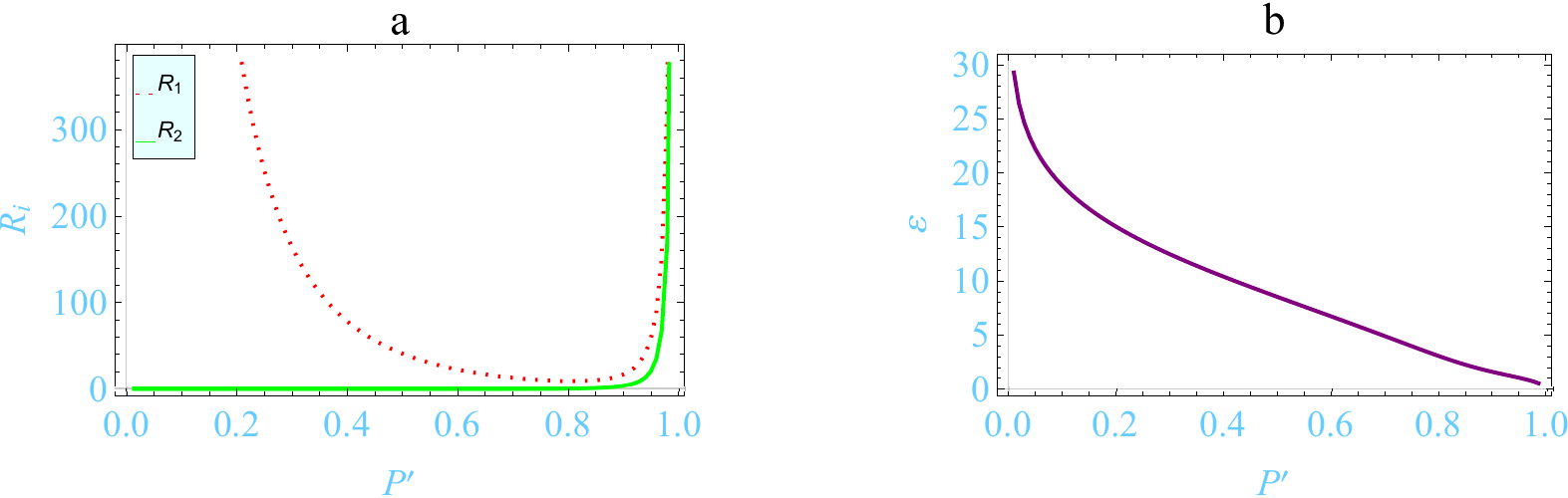}
   \caption{HEPM thermodynamical scalar curvatures of LBH and SBH along the coexistent curve of RN-AdS black holes (a) and HEPM thermodynamical scalar curvatures ratio between the LBH and SBH of  RN-AdS black holes (b) for $Q=1$.}
  \label{fignine}
\end{figure*}

\romannumeral2. As can be seen, $\varepsilon$ changes monotonically from non-zero to zero for dRGT massive black holes and holographic massive black holes. One may wonder that whether it suits for RN-AdS black holes. We have also calculated the HEPM scalar curvature for RN-AdS black holes as Fig.\ref{fignine}. There are indeed the same changing tendencies. One may predict that both the density difference and HEPM scalar curvature difference can be the order parameters for the phase transitions of AdS black holes.

\romannumeral3. The coexistent physics of four dimensional static spherically symmetric AdS black holes are explored in this paper. However, we believe that more valuable phenomena may be found for higher dimensional black holes, rotating black holes and black holes in grand canonical ensemble, which may contribute to our deeper understanding on thermodynamics of AdS black holes .

 \section*{Acknowledgements}
This work is supported by the National Natural Science Foundation of China (Grant Nos. 11235003, 11175019, and 11178007).

\bibliographystyle{elsarticle-num}
\bibliography{Notes}

\end{document}